\documentclass[runningheads]{llncs}
\newcount\DraftStatus  
\usepackage{comment}
\excludecomment{JournalOnly}  
\includecomment{ConferenceOnly}  
\excludecomment{TulipStyle}
\ifnum\DraftStatus=1
	\usepackage[draft,colorinlistoftodos,color=orange!30]{todonotes}
\else
	\usepackage[disable,colorinlistoftodos,color=blue!30]{todonotes}
\fi 
\makeatletter
 \providecommand\@dotsep{5}
 \def\listtodoname{List of Todos}
 \def\listoftodos{\@starttoc{tdo}\listtodoname}
 \makeatother
\usepackage{color}
\newcommand{\draftnote}[3]{ 
	\todo[author=#2,color=#1!30,size=\footnotesize]{\textsf{#3}}	}
\newcommand{\gangli}[1]{\draftnote{blue}{GLi:}{#1}}
\newcommand{\bbsong}[1]{\draftnote{red}{bbsong:}{#1}}
\newcommand{\kl}[1]{\draftnote{green}{kelvin:}{#1}}

\usepackage{graphicx}
\graphicspath{{./figures/}{./graphics/}{./graphics/logos/}}
\usepackage{pifont}
\usepackage{amsfonts,amssymb}
\usepackage{algorithm}
\usepackage{algorithmic}
\usepackage{breqn}
\usepackage{multirow}
\usepackage{psfrag}
\usepackage{url}
\usepackage{caption}
\usepackage{subfigure}
\usepackage[colorlinks,citecolor=blue]{hyperref}
\usepackage{cleveref}
\usepackage{booktabs}
\usepackage{rotating}
\usepackage{colortbl}
\usepackage{paralist}
\usepackage{setspace}
\usepackage{epstopdf}
\usepackage{nag}
\usepackage{microtype}
\usepackage{siunitx}
\usepackage{nicefrac}
\usepackage{fontawesome}
\usepackage{xcolor}
\usepackage{multicol}
\usepackage{wrapfig}
\usepackage{todonotes}
\usepackage{tablefootnote}
\usepackage{threeparttable}
\usepackage{cite}
\usepackage{lipsum}
\usepackage[english]{babel}
\usepackage[pangram]{blindtext}
\usepackage{tikz}
\usepackage[colorlinks,citecolor=blue]{hyperref}
\usetikzlibrary{fit,positioning,arrows.meta,shapes,arrows}
\begin{TulipStyle}
\usepackage[numbers]{natbib}
\usepackage{gitinfo2}
\usepackage{fancyhdr}

\usepackage{mathtools}
\usepackage{amsthm}
\theoremstyle{definition}

\theoremstyle{remark}

\numberwithin{equation}{section}

\newtheorem{theorem}{Theorem}

\newtheorem{definition}{Definition}

\hypersetup
{
    pdfauthor={\gitAuthorName},
    pdfsubject={TULIP Lab},
    pdftitle={},
    pdfkeywords={TULIP Lab, Data Science},
}

\end{TulipStyle}

\begin{document}
%
%
\title{Distributed Differentially Private \\Ranking Aggregation}
%
%
\author{
Baobao Song\inst{1}\orcidID{0000-0001-5630-5661}
\and
Qiujun Lan\inst{2}\orcidID{0000-0001-7523-9487}
\and
Yang Li\inst{3}\orcidID{0000-0002-1147-280X}
\and
Gang Li\inst{4}\orcidID{0000-0003-1583-641X}
}
%

\institute{
Hunan University, Changsha 410082, China 
\email{bbsong@tulip.academy}
\and
Hunan University, Changsha 410082, China 
\email{lanqiujun@hnu.edu.cn}
\and
The Australian National University, Canberra, ACT, 2600, Australia
\email{kelvin.li@anu.edu.au}
\and
Deakin University, Geelong, VIC 3216, Australia
\email{gang.li@deakin.edu.au}}

%
\maketitle              


\begin{abstract}
Ranking aggregation is commonly adopted in cooperative decision-making to 
assist in combining multiple rankings into a single representative. 
To protect the actual ranking of each individual, 
some privacy-preserving strategies,
such as differential privacy,
are often used. 
This, however, does not consider the scenario where the curator, 
who collects all rankings from individuals, is untrustworthy. 
This paper proposed a mechanism to solve the above situation 
using the \textsl{distribute differential privacy} framework. 
The proposed mechanism collects locally differential private rankings from individuals, 
then randomly permutes pairwise rankings using a shuffle model to further amplify the privacy protection. 
The final representative is produced by hierarchical rank aggregation. The mechanism was theoretically analysed and experimentally compared against existing methods, 
and demonstrated competitive results in both the output accuracy and privacy protection. 

\end{abstract}

\keywords{Ranking Aggregation, Distributed Differential Privacy, \texttt{HRA} Algorithm, Shuffle Model}%

\newenvironment{shrinkeq}[1]
{ \bgroup
  \addtolength\abovedisplayshortskip{#1}
  \addtolength\abovedisplayskip{#1}
  \addtolength\belowdisplayshortskip{#1}
  \addtolength\belowdisplayskip{#1}}
{\egroup\ignorespacesafterend}

\section{Introduction}\label{sec-intro}

Cooperative decision-making~\cite{meinhardt2012cooperative} is pervasive in business management,
because of its superiority in providing information from different aspects for better decision-making. 
As an essential step in cooperative decision-making, 
aggregation combines all individual preferences into a representative output. 
In daily life, 
individuals often rank all available alternatives to 
reveal the preference relation of multiple alternatives, 
hence \emph{ranking aggregation} has become essential for society, 
and many researchers focus on its two requirements,
which are hard to satisfy simultaneously: 
\emph{privacy} and \emph{utility}. 
Preference data in ranking has sensitive information, 
and the leaking of it may make individuals susceptible to coercion. 
\emph{Utility} represents whether the aggregation result stands for the majority preference.
Consequently, 
the ability to effectively aggregate private ranking into a representative result is important in ranking aggregation.
In the past few years, 
substantial research efforts have been devoted to ranking aggregation. 

Traditional anonymizing methods such
as anonymization hardly solves the problem.
For example, Hugo Awards 2015 incident~\cite{hugoawards} shows that 
the anonymized preferences could result in the re-identification of individuals 
because the adversary who has background knowledge are able to launch a \emph{linkage attack}. 
According to the weakness of the traditional anonymizing method, 
many researchers draw attention to \emph{differential privacy} (DP)~\cite{dwork2006differential}.

DP is an effective method to 
provide a rigorous privacy guarantee, 
and it can defend against various attacks, 
no matter how much background knowledge the adversary has.
As a lightweight methodology to protect privacy,
many current works address the ranking aggregation problems with DP. 
Shang et al.~\cite{shang2014application} designed a privacy-preserving rank aggregation algorithm,
and whatever the ranking rules, 
the algorithm adds noise on votes and returns the histogram of rankings.
Based on \texttt{Quicksort}~\cite{hoare1962quicksort},
Hay et al.~\cite{hay2017differentially} proposed three differentially private rank aggregation algorithms includes \texttt{P-SORT},
a pairwise comparison method about private ranking aggregation.
The benefit of using DP to protect individual sensitive information is that
the adversary is unlikely to obtain sensitive information by observing the releasing results.
In the meantime, 
the results have high availability.
Nevertheless, DP is not without limitations:
in real-world applications,
the curator may collude with the adversary to
leak some information before perturbing.

\emph{Local differential privacy} (LDP)~\cite{dwork2014algorithmic} alleviates this issue 
through adding noise locally and uploading noisy data to the untrusted curator.
Yan et al.~\cite{yan2020private} proposed \texttt{LDP-KwikSort} algorithm, 
and they use the number of queries $K$ to trade-off between utility and privacy. 
Besides privacy and utility, 
Wang et al.~\cite{wang2019aggregating} studied another property, soundness, 
and then proposed the weighted sampling mechanism and the additive mechanism to improve the ranking utility.
Unfortunately, 
LDP needs a large amount of data to achieve an acceptable utility.
Moreover, 
existing approaches that use pairwise comparison information to rank~\cite{yan2020private,hay2017differentially} share a common limitation:
they introduce additional errors by pivot random selection. 
In conclusion, 
two obstacles need to be overcome simultaneously. 
Firstly, 
the ranking algorithm needs to output an aggregation result with utility as high as possible. 
Secondly, 
in order to protect individuals' sensitive data, 
the untrusted curator should not receive the original raw preferences.
With the increased awareness of privacy protection,
many researchers are interested in \emph{distributed differential privacy} (DDP)~\cite{narayan2015distributed} to amplify the privacy.
DDP builds on LDP but further protects privacy using an intermediate node.
This may mitigate the problem of poor utility in LDP.
Besides, recently advances in ranking aggregation such as the algorithm HRA~\cite{ding2018new},
which takes advantage of pairwise comparisons to aggregate ranking,
and provides one way to eliminate the errors of random selecting pivot.
However, 
as this algorithm applies \emph{Borda count}~\cite{black1958theory} and pairwise comparison method, 
it costs too much privacy budget if it is directly combined with DP,
and results in very poor performance. 

In this paper, 
we propose an novel algorithm \texttt{DDP-Helnaksort} 
to meet the requirements of \emph{privacy} and \emph{utility} in ranking aggregation.
The contributions of this algorithm are two-fold: 
\begin{itemize}
    \item 
    \texttt{DDP-Helnaksort} employs a new ranking aggregation that avoids random pivot selection as in quicksort-based LDP methods. 
    Moreover, \texttt{Borda count} in \texttt{HRA} was replaced by a new method that scores the alternatives to reduce the noise effect caused by small privacy budget. 
    Experiments show that 
    it performs better than some pairwise comparison-based DP rank aggregations.
    
    \item  We firstly adopt DDP to deal with the ranking aggregation problem.
    This was achieved by combining LDP with a shuffle model~\cite{bittau2017prochlo} that  randomly permutes the preferences in order to amplify privacy before submitting rankings to the untrusted curator. This provides a stronger DP guarantee, which can be measured by calculating the amplification bound.


\end{itemize}
The rest of this paper is organized as follows. 
\Cref{sec-Preliminaries} provides the preliminary of ranking aggregation, 
differential privacy and shuffle model. 
\Cref{sec-Locally Private Hierarchical Rank Aggregation} 
presents the \texttt{DDP-Helnaksort} algorithm and gives the privacy guarantee. 
\Cref{sec-Experiment and Analysis} reports the comparison results 
with baseline algorithms and analyzes the effect of adjusting parameters. 
Final conclusions and future directions are shown in \cref{sec-Conclusions}.

\section{Preliminaries}\label{sec-Preliminaries}



\subsection{Ranking aggregation}\label{sec-Ranking aggregation}

\subsubsection{Conception and Measurement}\label{sec-Conception and Measurement}


In a ranking scenario, 
an agent $u$ is asked to rank a set $A=\{a_1,a_2,...,a_m\}$ of alternatives 
and to provide the order of preferences,
denoted by $p_u=[x_1,x_2,...,x_m]$, 
where $x_i$ is the ranking index of $a_i$, 
and $x_i=1$ means that agent $u$'s favourite alternative is $a_i$.
A curator then collects the order from each agent 
and uses a ranking aggregation algorithm to output a representative ranking $R$ based on $\{p_u\}$.
In this paper, 
$a_i \succ a_j$ means that $a_i$ is preferred than $a_j$.



Ranking aggregation aims to find the most representative ranking $R^*$. 
\emph{Kenmeny optimal aggregation} (KOA)~\cite{dwork2001rank} is used to 
find $R^*$ by minimising the \emph{average Kendall Tau distance} $\overline{K}$.
\emph{Kendall tau} distance~\cite{kendall1948rank} measures the distance between two rankings 
by counting the number of inconsistent pairs among all pairs of alternatives:
$K(R,p_u)=\frac{1}{\binom{m}{2}}\sum_{i\neq j,i,j\in[m]} \kappa_{ij}(R,p_u)$, 
$\kappa_{ij}(R,p_u)$ is $1$ when the pair $a_i$ and $a_j$ is ordered differently in rankings $R$ and $p_u$,
otherwise it is $0$.
The average \emph{Kendall tau distance} 
is then computed over the rankings $\{p_u\}$ from all agents:
$\overline{K}(R,p_u)=\frac{1}{n}\sum_{u\in[n]} K(R,p_u)$.

\subsubsection{Hierarchical Ranking Aggregation }\label{sec-Hierarchical Ranking Aggregation}


 
\texttt{Hierarchical ranking aggregation} (HRA)~\cite{ding2018new} algorithm 
can consolidate all agents' rankings into a total order. 
It is a recursive process like \texttt{Quicksort}, 
but does not rank alternatives based on the random pivot selections as in \texttt{Quicksort}, 
which is likely to reduce the utility of a private ranking 
besides the impact of additive noise such as in DP.  

Given $m$ ranking alternatives, 
the \texttt{HRA} algorithm first computes an $m$ by $m$ \emph{pairwise comparison matrix} (PCM) $M$, 
where each entry $M(i,j)=\frac{1}{n}\sum_{u=1}^n l_{ij}^u$ is a comparison score 
for $(a_i,a_j)$ over all $n$ rankings. 
$l_{ij}^u=1$ and $0$ when $a_i \succ a_j$ and $a_j \succ a_i$ respectively and $0.5$ otherwise. 
Then, the algorithm computes an $m$ by $m$ \emph{pairwise preference relation} (PPR) matrix $D$. 
Each entry $D(i,j)=1$ or $0$ 
when $M(i,j)$ is greater or smaller than $M(i,j)$ respectively, 
and $0.5$ for equality.
Third, 
every alternative is allocated to a different level 
according to its score $L(i)$ that is the row sum of $D(i,j)$. 
As multiple alternatives can be allocated at the same level, 
they are further compared and allocated into sublevels 
using a sub-PCM that only includes the rankings of the corresponding alternatives,.
If alternatives have the same score in sublevel, 
\texttt{Borda count} is used to select a winner. 
Finally, 
the algorithm finishes when each level contains only one alternative.

\subsection{Differential privacy}\label{sec-Differential privacy}

Differential privacy (DP)~\cite{dwork2006differential} is a privacy protection model 
that adds calibrated noise to query outputs to ensure an adversary 
having negligible chance of guessing the sensitive information in a database. 
Formally, DP can be defined as: 
\begin{definition}\label{def:def1}
\textbf{$(\epsilon,\delta)$-Differential Privacy}. 
A random algorithm $\mathcal{M}$ provides $(\epsilon,\delta)$-differential privacy 
if for any two datasets $D$ and $D'$ that differ in at most a single record, 
and for all outputs $A \in Range(O)$: $Pr[\mathcal{M}(D)\in O)]\leq e^\epsilon Pr[\mathcal{M}(D')\in O)]+\delta$.
\end{definition}

The parameter $\epsilon$ is defined as the privacy budget,
which controls the privacy guarantee level of the mechanism.
Another parameter $\delta$ is responsible for the probability that $\epsilon$ does not hold.
DP assumes that there is a trusted curator, 
but in reality, 
the adversary has possibility to collect information from the curator. 
Hence,
the local differential privacy (LDP)~\cite{dwork2014algorithmic} has been utilized. 
In LDP model, each agent uses an algorithm $\mathcal{M}$ to perturb data locally and then upload the noisy one to the untrusted curator.
The definition of LDP is as follows:
\begin{definition}\label{def:def3}
\textbf{$(\epsilon,\delta)$-Local Differential Privacy}. 
A local algorithm $\mathcal{M}$ provides $(\epsilon,\delta)$-local differential privacy if for any two value $x$ and $x'$, and for every output $y$:  $Pr[\mathcal{M}(x=y)]\leq e^\epsilon Pr[\mathcal{M}(x'= y)]+\delta$.
\end{definition}

Although LDP solves the problem that the curator may disclose information,
it requires a huge amount of data to achieve a satisfactory utility \cite{bittau2017prochlo}. 
Based on LDP, 
\emph{distributed differential privacy} (DDP) \cite{narayan2015distributed} can improve the data utility.
In DDP, every agent adds noise locally, 
and uploads the data to a trusted intermediate node to protect privacy further,
finally sends the results to the curator. 
On the one hand, 
we do not need to worry about the privacy leakage from the curator in DDP. 
On the other hand, 
it has a higher utility than LDP.
In this paper, 
we apply DDP model to aggregate ranking.
And we use Gaussian mechanism~\cite{dwork2014algorithmic} to perturb preferences.
An application of Gaussian mechanism satisfies $(\epsilon,\delta)$-DP, 
if variables drawn from the Gaussian distribution with $\mu=0$ and 
$\sigma=\frac{\Delta_g f \sqrt{2\log(\frac{1.25}{\delta})}}{\epsilon}$, 
where $\Delta_g f$ is the global sensitivity of the function $f$.

\subsection{Shuffle model}\label{sec-Shuffle model}

Shuffle model can be used in intermediate nodes to realise DDP, 
and the protocol $P$ was proposed in \cite{bittau2017prochlo}. 
The protocol has three components: a randomizer $\mathcal{R}$,
a shuffler $\mathcal{S}$ and an analyzer $\mathcal{A}$. 
First, $\mathcal{R}$ applies LDP to perturb data to get $(\epsilon,\delta)$ protection. 
Then, $\mathcal{S}$ chooses a random permutation $\pi$ to shuffle the data, 
and cut the connection between the outputs and their sources. 
Finally, 
$\mathcal{A}$ analyses the data and gets the query result. 
The shuffle step can amplify the privacy, 
and the following theorem~\cite{cheu2019distributed} quantifies the amplification bound of shuffling:
\begin{theorem}\label{the:the4}
If every agent sends a message to the shuffle model, 
and the randomizer $R$ satisfies $(\epsilon+\ln{n},\delta)$-local differential privacy, 
then the protocol $P=(R,S,A)$ satisfies both $(\epsilon,\delta)$-differential privacy 
and $(\epsilon+\ln{n},\delta)$-local differential privacy, 
where $\epsilon'$ is smaller than $\epsilon$,
and $n$ is the  number of agents.
\end{theorem}

\section{Ranking Aggregation Algorithm Under DDP}
\label{sec-Locally Private Hierarchical Rank Aggregation}


In this section, 
we propose an algorithm \texttt{DDP-Helnaksort} to solve the private ranking aggregation problem.
It can be formalised as follows: 
given $m$ alternatives to be ranked by $n$ agents, 
a curator need to present a final ranking that represents most agents' preferences. 
In addition, 
each agent $u$'s ranking $p_u$ must not reveal to the curator his true preferences over the alternatives.

\begin{figure}[!htbp]
    \centering
    \setlength{\belowcaptionskip}{-0.3cm}
    \includegraphics[width=\textwidth]{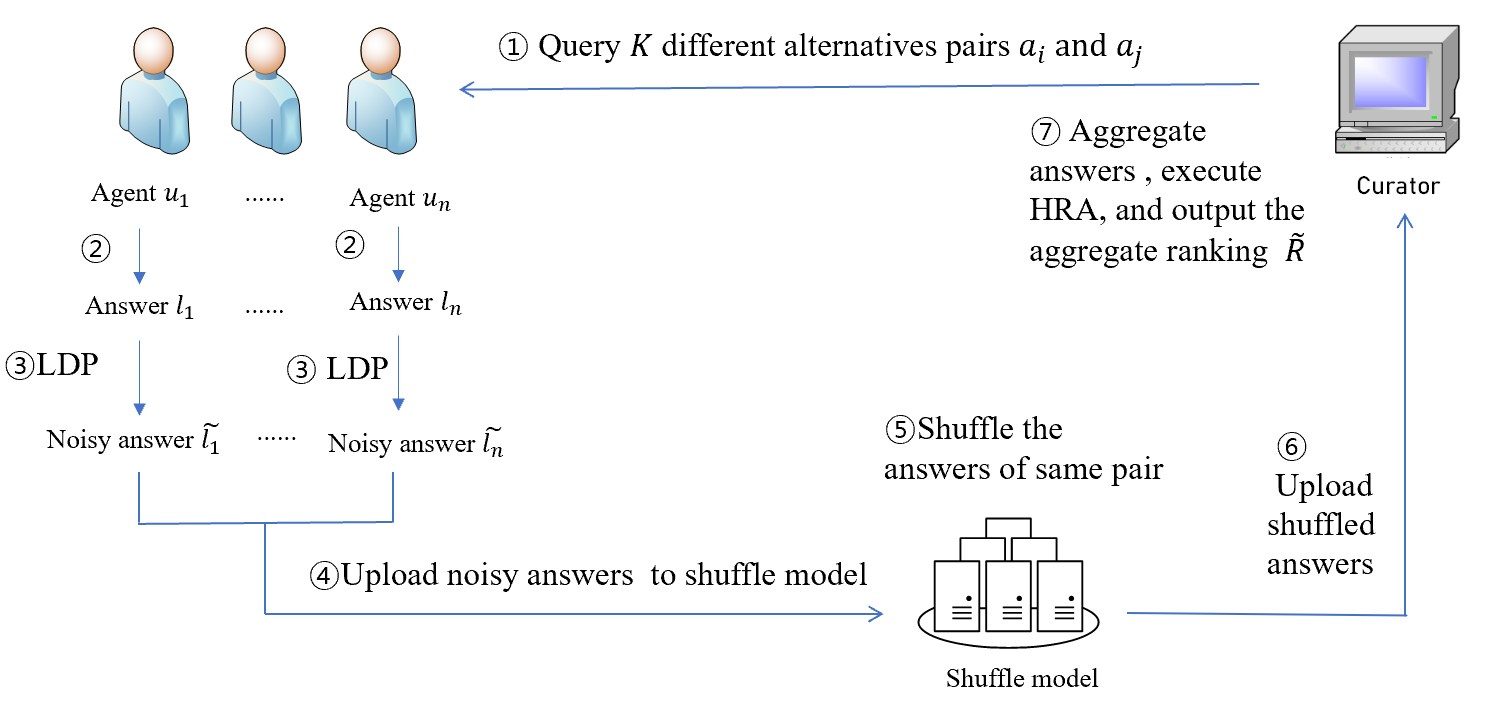}
    \caption{Overview of \texttt{DDP-Helnaksort}}
    \label{fig:fig7}
\end{figure}

The \texttt{DDP-Helnaksort} algorithm consists of three steps, as shown in \cref{fig:fig7}. 
These steps are discussed respectively in \cref{sec-Ranking Preference Collection},  
\cref{sec-Shuffling} and \cref{sec-Ranking Aggregation}. 
The first step (\ding{172}-\ding{175}) is ranking preference collection, 
in which each agent, before submitting the answers to the curator, 
adds the Gaussian noise to the rank of each pair $(a_i,a_j)$ that being queried. 
The second step (\ding{176}-\ding{177}) is a shuffling process, 
which collects the ranking of $(a_i,a_j)$ from the corresponding agents that answered the query, 
in order to further reduce the risk of privacy breach.
The third step (\ding{178}) aggregates to generate a final ranking of all $m$ alternatives.
\vspace{-0.2cm} 

\subsection{Ranking Preference Collection}
\label{sec-Ranking Preference Collection}
The first step collects $K$ private pairwise rankings from each agent, where $K$ is an input parameter. 
A larger $K$ leads to a more accurate aggregated ranking, because each pair $(a_i,a_j)$ will be answered by more agents. The drawback is the partition of the privacy budget into a tiny piece for each query, which results in adding large noise that diminishes the utility. A smaller $K$ can guarantee the utility, but the curator may end up with a less representative final ranking. We explore the optimal $K$ in \cref{sec-Performance of DDP-Helnaksort}. The ranking preference collection step is shown in Algorithm \cref{alg:alg1}.
$l_{ij} \leftarrow p_u$ represents the preference in agent $u$'s ranking of a randomly selected pair $(a_{i},a_{j})$. 
This algorithm uses the Gaussian mechanism for noise addition (other mechanisms can be used too).
\begin{algorithm}
\setstretch{0.93}
\renewcommand{\algorithmicrequire}{\textbf{Input:}}
	\renewcommand{\algorithmicensure}{\textbf{Output:}}
	\caption{Ranking Preference Collection}
	\label{alg:alg1}
	\begin{algorithmic}[1]
		\REQUIRE Agent $u$'s ranking $p_u$, $K$ queries, privacy parameter $\epsilon$ and $\delta$
		\ENSURE Private pairwise preferences $Q$
		\STATE $Q_u=\emptyset$
		\FOR{$k \in [K]$}
		    \STATE $l_{ij} \leftarrow p_u$
		    \STATE $\widetilde{l}_{ij}=l_{ij}+Gau(\frac{K\Delta f \sqrt{2\ln {\frac{1.25}{\delta}}}}{\epsilon})$
    	    \IF{$\widetilde{l}_{ij} > 0.5$}
    		    \STATE$\widetilde{l}_{ij}=1$
    		\ELSE
    		    \STATE$\widetilde{l}_{ij}=0$
    		\ENDIF
    		\STATE $Q_u=Q_u \cup \{\widetilde{l}_{ij}\}$ 
        \ENDFOR
	\end{algorithmic}  
\end{algorithm}

\subsection{Shuffling}\label{sec-Shuffling}


Shuffling before aggregation can amplify privacy without affecting the output utility. 
In \texttt{DDP-Helnaksort}, 
each pair $(a_i,a_j)$'s answers from the corresponding agents are collected and randomly permuted at an intermediate node, so that when the private rankings are submitted, 
the curator is unable to guess the source of an answer with a non-negligible probability.
The shuffle model finally provides a protection of DP with a smaller $\epsilon$, 
which is further discussed in \cref{sec-Privacy Guarantee}.

\subsection{Ranking Aggregation}\label{sec-Ranking Aggregation}

Once all the private rankings are submitted, the \texttt{DDP-Helnaksort} algorithm goes into the final stage, ranking aggregation. This step is based on the \texttt{HRA} algorithm, but with a different fallback to sort equal alternatives in sublevels in order to reduce the noise effect. The algorithm is shown in \cref{alg:alg2}. 
$M$ is the number of alternatives in the unsorted sublevel, 
and $C_{a_{i}a_{j}}$ is the number of agents who voted $a_i \succ a_j$.
The method uses pairwise preference to calculate a score for $a_j$
\begin{shrinkeq}{-1ex}
\begin{equation}
\label{equ:ec3}
    C_{a_{j}}=\sum_{j \in [M]} (C_{a_{j}a_{i}}-C_{a_{i}a_{j}}),
\end{equation}
\end{shrinkeq}
hence avoids splitting some privacy budgets as in \emph{Borda count}. 

This \texttt{RA}(ranking aggregation) algorithm mainly adopts a separate-layer ranking thought to generate the aggregation ranking, 
which uses the information about $C_{a_{i}a_{j}}$ and $C_{a_{j}a_{i}}$. 
The calculations of PCM and PPR matrix happen at $Line$ \num{6}-\num{8} and $Line$ \num{9}, respectively. 
After that, we can count the scores of every alternative in $M$ ($Line$ \num{10}-\num{12}). 
And if the scores are same in two rounds, 
we calculate the $C_{a_{j}}$, 
and then put the highest one in a high level and others at a low level to do the next round ($Line$ \num{13}-\num{16}). 
The algorithm iterates until $M=1$ in each level, 
and finally the aggregated ranking $\widetilde{R}$ is generated ($Line$ \num{17}-\num{20}).


\begin{algorithm}[htb]
    \setstretch{0.93}
	\renewcommand{\algorithmicrequire}{\textbf{Input:}}
	\renewcommand{\algorithmicensure}{\textbf{Output:}}
	\caption{RA}
	\label{alg:alg2}
	\begin{algorithmic}[1]
		\REQUIRE Agents pairwise aggregation  $C_{a_{i}a_{j}}$ and $C_{a_{j}a_{i}}$
		\ENSURE Aggregate ranking $\widetilde{R}$
		\STATE $M$=number of alternatives needed to rank
		\STATE $L=[0]*M$
		\IF{$M=1$}
		    \RETURN
		\ENDIF
		\FOR{each $i$, $j \in [m]$}
		    \STATE Calculate $PCM(i,j) =\frac{C_{a_{i}a_{j}}}{C_{a_{i}a_{j}}+C_{a_{j}a_{i}}}$
		\ENDFOR
		\STATE Calculate $PPR$ according to $PCM$
		\FOR{$j, i \in [M]$}
	    \STATE Calculate alternatives' level score $L(j)+=PPR(i,j)$
	    \ENDFOR
	    \IF{$L(1)=L(2)=...=L(M)$}
	        \STATE put the $C_{a_{j}}$ winner into a high-ranking level and others into a low level 
	    \ENDIF
	    \FOR{$l=1$ to the number of different levels}
	        \STATE ranking of $l$-th level=$\texttt{RA}$ (input ranking about the alternatives in $l$-th level)
	    \ENDFOR
	    \STATE Rank the alternatives according to their levels to get aggregate ranking $\widetilde{R}$
	\end{algorithmic}  
\end{algorithm}


\subsection{Privacy Guarantee}\label{sec-Privacy Guarantee}


\begin{theorem}\label{the:the2}
$\texttt{DDP-Helnaksort}$ satisfies $(\epsilon,\delta)$-local differential privacy 
and  $(\epsilon-\ln{\frac{n}{\binom{m}{2}}}, \delta)$-differential privacy when $K=1$.
\end{theorem}

\begin{proof}
In the ranking preference collection phase,
Gaussian mechanism is used to add noise into every agent's answers. Because are $K$ rounds,
$\epsilon_k=\frac{\epsilon}{K}$ in each round.
In Gaussian mechanism, 
we set 
\begin{equation}\label{equ:ec10}
\delta=\frac{\Delta_g f \sqrt{2\ln {\frac{1.25}{\delta}}}}{\epsilon_k}=\frac{K\Delta_g f \sqrt{2\ln {\frac{1.25}{\delta}}}}{\epsilon}
\end{equation}
And \texttt{DDP-Helnaksort} executes the post-processing procedure after applying Gaussian mechanism, 
hence it satisfies $(\epsilon,\delta)-LDP$. 
Besides, 
$K=1$ means that every agent answers once and 
uploads a single message (latter experiments confirm the algorithm utility is the highest when $K=1$). 
In the shuffling phase, there are ${\binom{m}{2}}$ pairs of alternatives, so the number of same pair and the size of set $S$ in shuffle model is 
\begin{equation}\label{equ:ec10}
n'=\frac{n}{\binom{m}{2}}
\end{equation}
Therefore, 
by using \cref{the:the4}, 
the algorithm \texttt{DDP-Helnaksort} 
satisfies $(\epsilon-\ln{\frac{n}{\binom{m}{2}}}, \delta)$-DP when $K=1$.
\end{proof}

\section{Experiments}
\label{sec-Experiments}

In this section, 
we evaluate the performance of \texttt{DDP-Helnaksort},
and compare it with benchmark methods on both real and synthetic datasets. 
All algorithms were implemented in Python and executed 300 times to get the result.

\subsection{Experiment Settings}\label{sec-Experiment Settings}

\subsubsection{Datasets}\label{sec-Datasets}

The experiments were conducted on synthetic datasets
and a real-world dataset \textbf{TurkDots}~\cite{mao2013better}.
By using R package \texttt{PerMallows} \num{1.13}, 
we obtained four synthetic datasets with $n \in\{100,1000,2500,5000\}$, 
$\theta=0.25$, and $m=15$ from Mallows model~\cite{mallows1957non}. 
The dispersion parameter $\theta$ represents the distance between the generated ranking and ground truth ranking. 
The generated ranking is closer to the ground truth ranking 
when $\theta$ is larger. 
\textbf{TurkDots} is from \emph{Amazon Mechanical Turk},
and it contains $m=4$ alternatives rankings. 


\subsubsection{Baseline Algorithms}\label{sec-Compared Algorithms}

\begin{itemize}
    \item \texttt{LDP-Kwiksort}~\cite{yan2020private}.
    It has $K$ rounds' interactions between every agent 
    and the untrusted curator. 
    In each round, 
    the curator random selects paired alternatives to ask agents preference and receives noisy answers from agents (queries to an agent are not the same),
    then uses the \texttt{Kwiksort} algorithm to get the aggregate ranking. 
    Its utility is the highest when $K=1$.
    \item \texttt{LDP-Quicksort}. 
    Compared with \texttt{LDP-Kwiksort},
    it only differs in when a new pivot random chosen in \texttt{Quicksort}, 
    the curator will query the preference between the pivot and other alternatives.
    This setting is only to collect preference used in \texttt{Quicksort}, 
    and avoid the waste of privacy budget for other pairs.
    $K$ in this algorithm represents the times of the agent's answers.
    Finally, when the \texttt{Quicksort} algorithm is finished, 
    the curator gets an aggregated ranking. 
\end{itemize}

\subsubsection{Utility Metric - \textbf{Average Kendall tau distance}}
\label{sec-Utility Metric}

We use the \emph{average Kendall tau distance} to 
measure the accuracy of the aggregated ranking. 
The larger the \emph{average Kendall tau distance}, 
the worse the algorithm performance. 
We normalise this distance by $m(m-1)/2$ 
because we can directly compare it with different number of alternatives. 
Hence, 
the \emph{average Kendall tau distance} can be calculated as 
$\overline{K}(R,R_u)=\frac{2}{nm(m-1)}\sum_{u\in N}K(R,R_u)$.

\subsection{Performance of \texttt{DDP-Helnaksort}}
\label{sec-Performance of DDP-Helnaksort}

\subsubsection{Comparison between \texttt{DDP-Helnaksort} and Baseline algorithms}
\label{sec-Comparison between DDP-Helnaksort and other algorithms}

\begin{ConferenceOnly}
We ran three algorithms \texttt{LDP-Quicksort},
\texttt{LDP-Kwiksort}, \texttt{DDP-Helnaksort} with Gaussian noise.
Here $\epsilon$ is the parameter in LDP. 
We set $K \in \{1,m,max\}$ to observe the performance of different algorithms in different $K$, 
and $m$ is the number of alternatives.
When $K=max$, 
the maximum value of $K$ in \texttt{LDP-Kwiksort} and \texttt{DDP-Helnaksort} is $\binom{m}{2}$, 
but in \texttt{LDP-Quicksort}, 
the value is according to the chosen pivot, 
and it is $(m-1)\log m$ in general. 
We did the experiment on \textbf{TurkDots} with $n=100$.
With $\epsilon=1$, $\delta=10^{-4}$ in local differential privacy,
the \emph{average Kendall tau distance} of \texttt{LDP-Quicksort}, 
\texttt{LDP-Kwiksort} and \texttt{DDP-Helnaksort} are shown in \cref{fig:fig2}.

\begin{figure}[h]
    \vspace{-0.4cm}
    \centering
    \setlength{\belowcaptionskip}{-0.5cm}
    \includegraphics[scale=0.27]{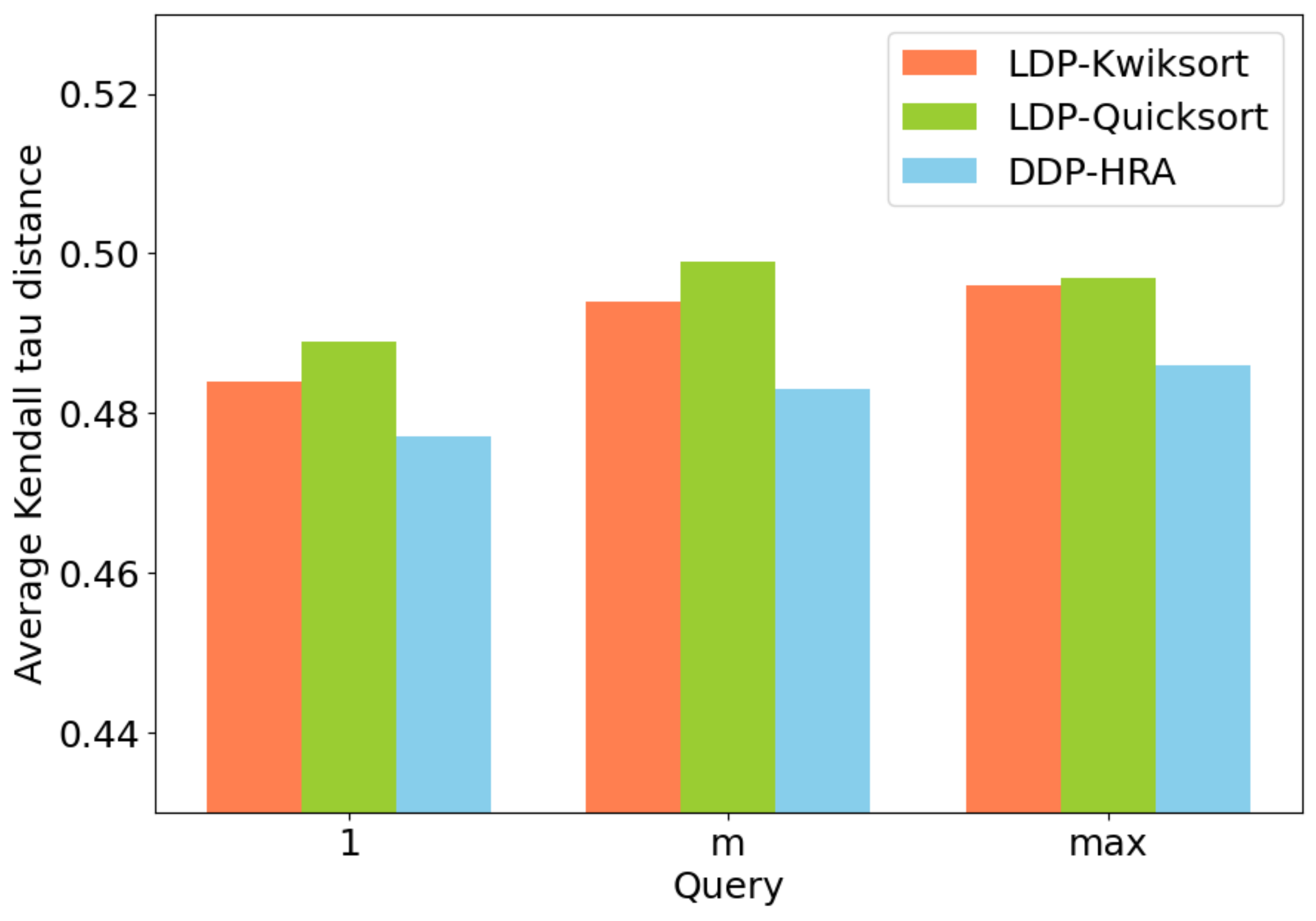}
    \caption{Comparison of algorithms according to \emph{average Kendall tau distance} 
    on \textbf{TurkDots} across different K}
    \label{fig:fig2}
\end{figure}

The results in \cref{fig:fig2} proves our algorithm outperforms others across different $K$.
When we add the same scale of noise to these algorithms, 
the \emph{average Kendall tau distance} of \texttt{DDP-Helnaksort} 
is the shortest.
The cause is when adding same scale of noise,
\texttt{DDP-Helnaksort} uses more pairwise alternatives' information
(the comparison information provided by pairwise comparisons) to rank,
which leads to a more accuracy result.
Besides, it keeps away from the error of pivot random selection, 
which can not be avoided by the other algorithms.
\end{ConferenceOnly}

\subsubsection{Impact of Query Amount to every agent}\label{sec-Impact of Query Amount}

\begin{figure}
\vspace{-0.5cm}
\centering 
\setlength{\abovecaptionskip}{-0.05cm}
\setlength{\belowcaptionskip}{-0.5cm}
\subfigcapskip=-3pt
\subfigure[]{ 
\label{fig:subfig:a} 
\includegraphics[scale=0.32]{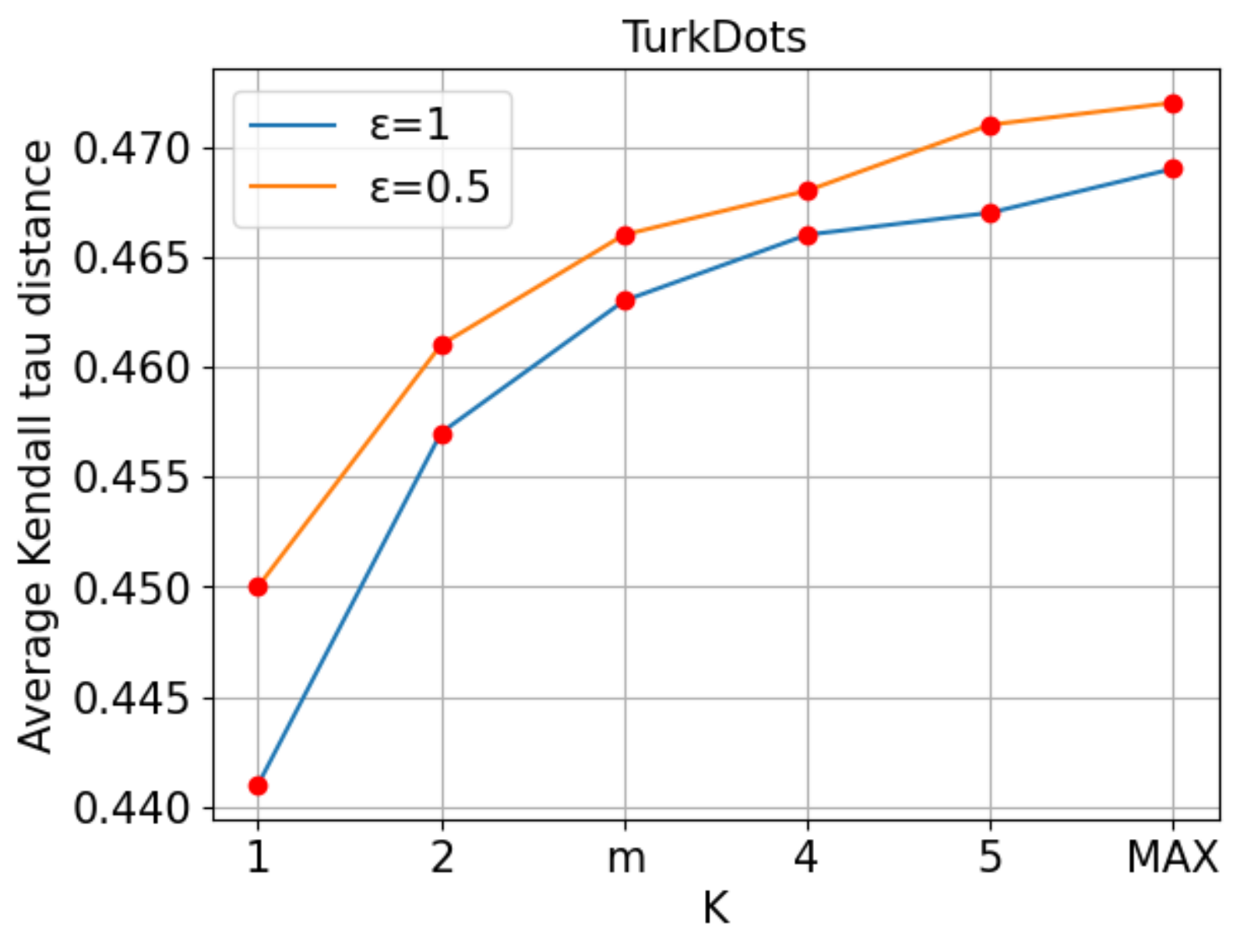}} 
\hspace{0.1in} 
\subfigure[]{ 
\label{fig:subfig:b} 
\includegraphics[scale=0.32]{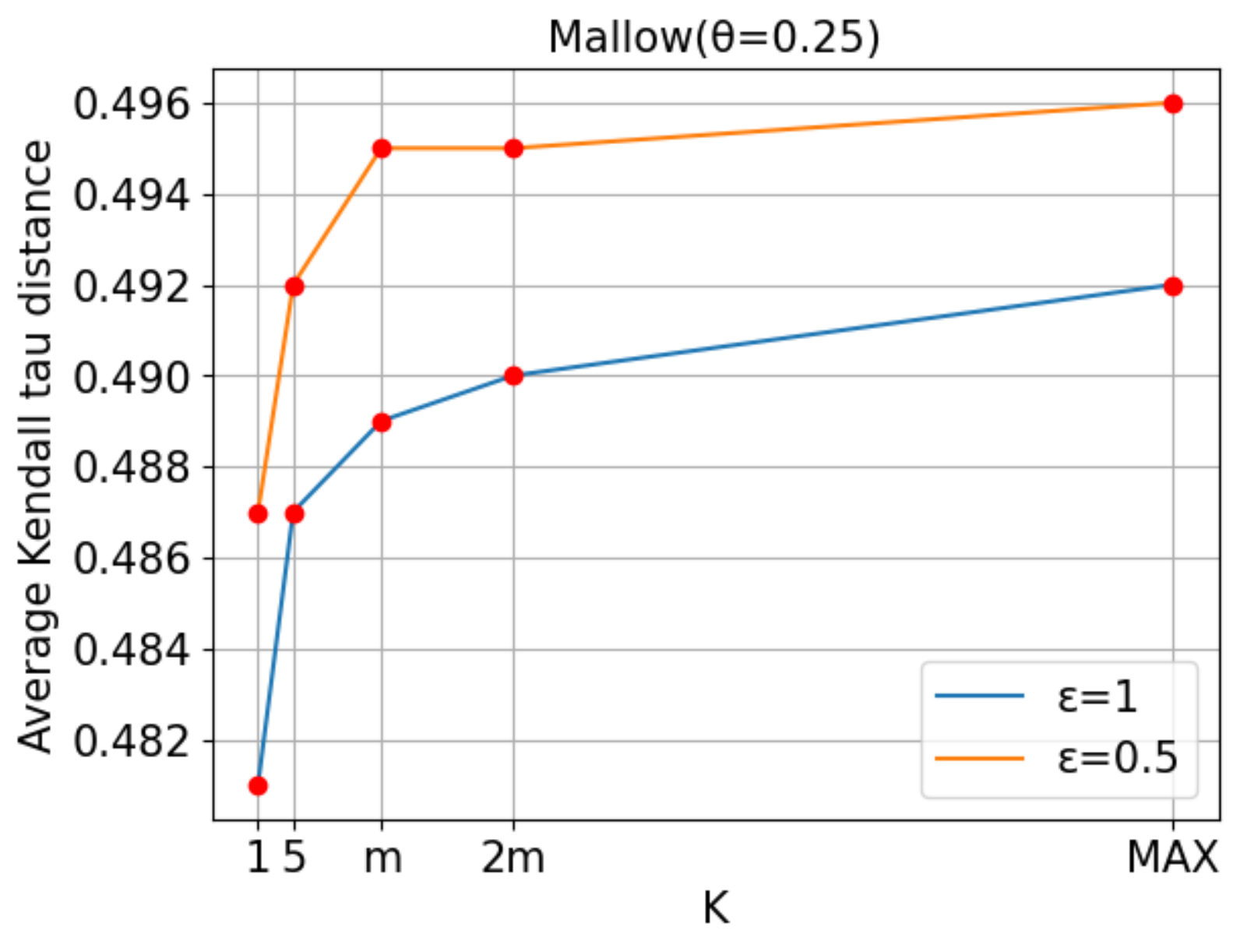}} 
\caption{Performance of \texttt{DDP-Helnaksort}: 
\emph{Average Kendall tau distance} on \textbf{TurkDots} (a) and a synthetic dataset 
(b) across different K when $\delta=10^{-4}$, $\epsilon \in \{0.5, 1\}$} 
\label{fig:fig3} 
\end{figure}

Different number of the queries has different ranking aggregation results. 
More information can be obtained when increasing the number of queries, 
but at the same time, 
the privacy budget of each round becomes smaller, 
and the larger noise is added to every answer. 
In order to get the best performance with the best $K$, 
we ran \texttt{DDP-Helnaksort} on dataset \textbf{TurkDots} 
and the synthetic dataset with $100$ agents.
We set $\delta=10^{-4}$, $\epsilon \in \{0.5, 1\}$ 
(this $\epsilon$ is the parameter of DP,  
also means that it is the amplification result of local randomizer,
and $\epsilon$ in following experiment is the same)
as well as varying the number of queries $K$ to 
observe the performance of \texttt{DDP-Helnaksort}. 
The results are shown in \cref{fig:fig3}.

It is apparent that as the decreasing of $K$, 
the performance of \texttt{DDP-Helnaksort} is better. 
The \emph{average Kendall tau distance} reaches the minimum when $K=1$. 
This experiment result is the same as~ \cite{yan2020private}, 
which reveals the best performance is achieved when $K=\frac{\epsilon}{2}$. 
The reason of this phenomenon is large $K$ leads to a  small $\epsilon$ in each round, 
and large scale of noise has a great impact on results.
Although some information about agents' preferences is lost when $K$ is small, 
a small noise is added to each answer, 
and the impact is smaller than large noise with more information. 
The result also implies if we want to further improve performance of the algorithm, 
we can do some works about handling $\epsilon$ 
such as implementation of personalised differential privacy
which can release some needless privacy. 

\subsubsection{Ablation Study: Impact of shuffle model and privacy budget}\label{sec-Impact of shuffle model}

As seen in \cref{sec-Privacy Guarantee},
shuffle model turns LDP to DDP and amplifies the privacy. 
When every agent gives his noisy answers,
a shuffling mechanism used before aggregation can offer another protection. 
After using the shuffle model, 
the algorithm satisfies DP with a smaller $\epsilon$ than before. 
In order to demonstrate the privacy amplification of shuffling,
we compared the algorithm with and without shuffle model in a same $\epsilon$. 
Besides,
$\epsilon$ reflects the level of privacy protection of every agent.
We varied $\epsilon$ to observe the changes in \emph{average Kendall tau distance}.
We set $k=1$, and other experimental setup is unchanged.

\begin{figure} 
\setlength{\abovecaptionskip}{-0.05cm}
\setlength{\belowcaptionskip}{-0.5cm}
\subfigcapskip=-3pt
\centering 
\subfigure[]{ 
\label{fig:subfig:a} 
\includegraphics[scale=0.32]{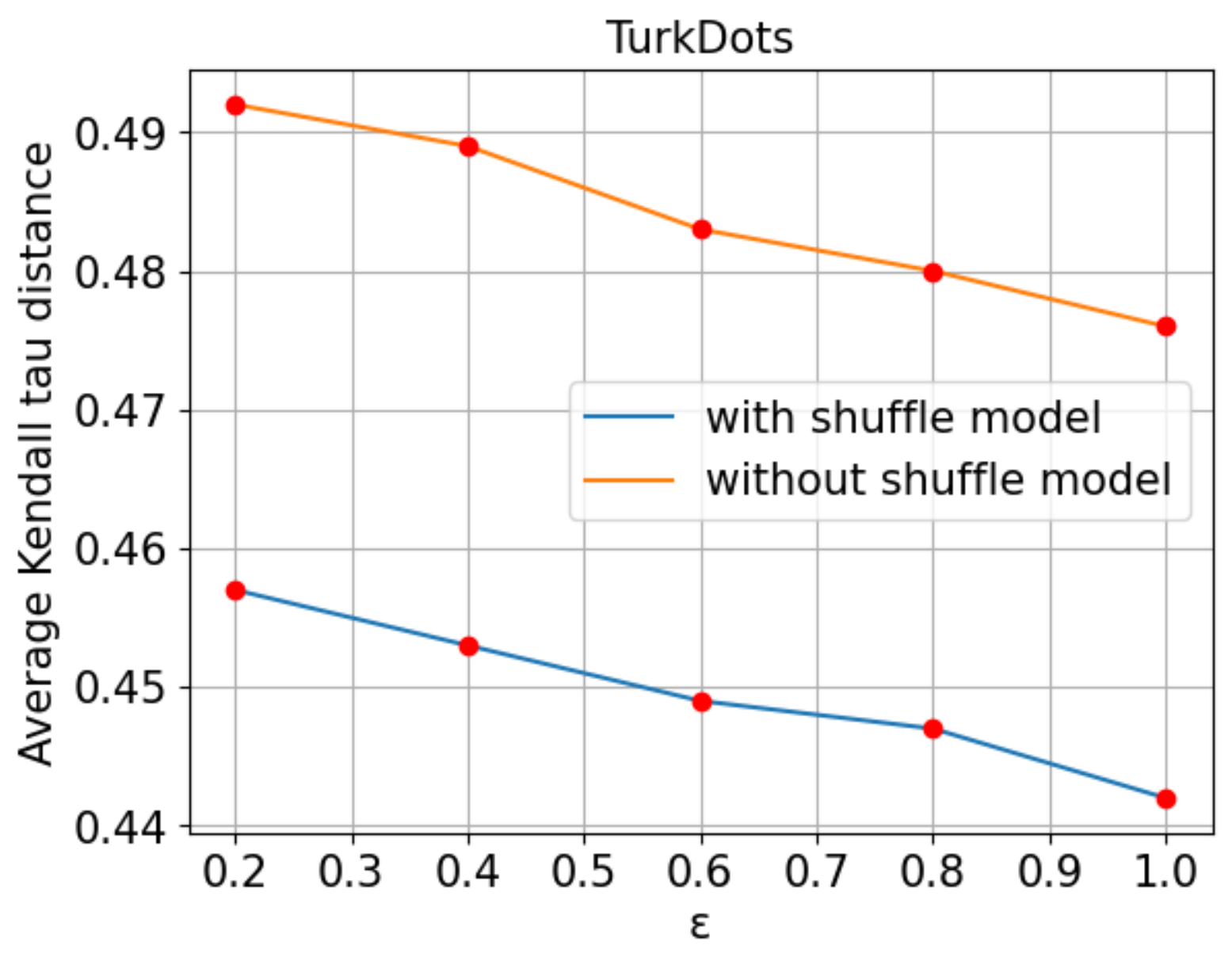}} 
\hspace{0.1in} 
\subfigure[]{ 
\label{fig:subfig:b} 
\includegraphics[scale=0.32]{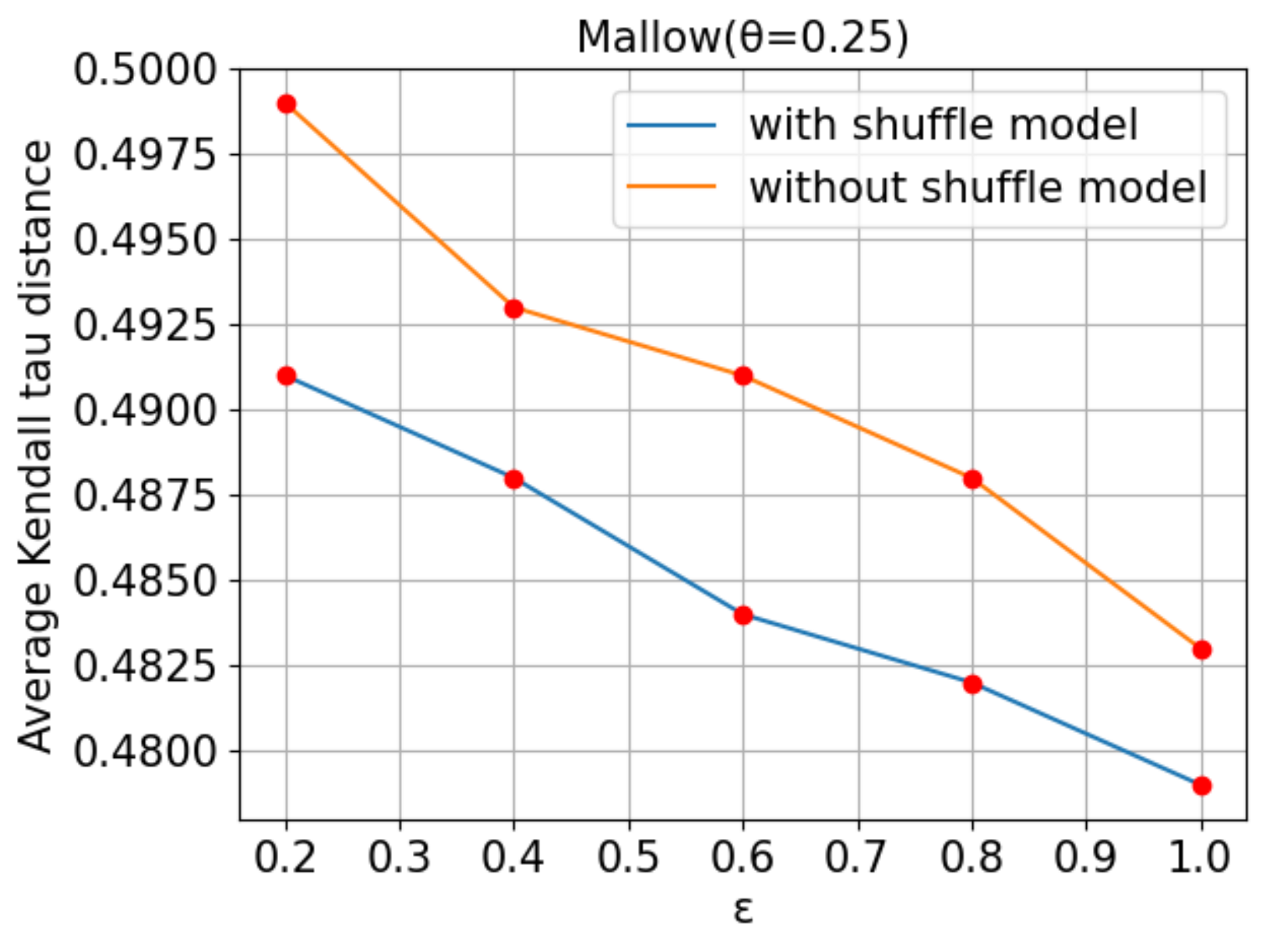}} 
\caption{Comparison of DDP-Helnaksort with and without shuffle model according to \emph{average Kendall tau distance} on \textbf{TurkDots} (a) and a synthetic dataset (b) across different $\epsilon$ when $K=1$ and $n=100$} 
\label{fig:fig6} 
\end{figure}

We can conclude from \cref{fig:fig6} that 
adding the shuffle step results in a better utility.
The reason is that shuffling is equivalent to adding another noise on data. 
Consequently, 
when we compared the algorithm with and without shuffling at a certain $\epsilon$,
the second one has a large $\epsilon$ locally, so it perturbs less on data and performs better.  
In \cref{fig:fig6}, 
the distance average increases more in \textbf{TurkDots} 
than the synthetic dataset from with shuffling to without shuffling,
and this mainly relates to different number of alternatives $m$. 
The synthetic dataset has more alternatives than \textbf{TurkDots}, 
thus the synthetic dataset has more alternative pairs
and it has fewer collected preferences about a certain pair.
Therefore, 
the shuffle model offers a smaller amplification on the synthetic dataset. 
This phenomenon is consistent with the \cref{the:the2} that the amplification bound is proportional to the amount of data about a certain pair. 
Moreover, 
when decreasing the privacy budget, 
the \emph{average Kendall tau distance} increases due to large scale of noise, 
which make the final aggregation ranking further to the representative ranking.
Furthermore, 
in DDP, 
we can choose alternative methods,
such as some cryptography tools,
to amplify the privacy.

\section{Conclusions}\label{sec-Conclusions}
In order to improve the utility of private ranking aggregation,
we proposed a new algorithm \texttt{DDP-Helnaksort},
which avoids the issue of  random pivot selection 
which appears in other private ranking algorithm using the pairwise method.
We designed a new method to give alternatives' score according to preference in pairs, 
which can save some privacy budget and lead to a higher utility.
Experimental results indicate that
our algorithm achieves a better performance.
Besides,
We're first applying the DDP mechanism shuffle model to amplify the privacy.
Theoretical analysis of amplification bound of shuffle model 
and experimental results all confirm that the shuffle model is valid.

In the future, 
we will further improve the ranking utility, 
such as using some cryptography tools. 
Besides, 
this algorithm can be further optimised 
if it could apply personalised DP, 
which can release some redundant privacy budget to achieve a higher utility.


\bibliographystyle{IEEEtran}



\end{document}